\documentclass[lettersize,journal]{IEEEtran}
\usepackage{amsmath,amsfonts}
\usepackage{algorithmic}
\usepackage{algorithm}
\usepackage{array}
\usepackage[caption=false,font=normalsize,labelfont=sf,textfont=sf]{subfig}
\usepackage{textcomp}
\usepackage{stfloats}
\usepackage{url}
\usepackage{verbatim}
\usepackage{graphicx}
\usepackage{cite}
\usepackage{multirow} 
\usepackage{booktabs}
\usepackage{xcolor}
\usepackage{caption}
\usepackage{hyperref}
\begin{document}

\title{Sync-TVA: A Graph-Attention Framework for Multimodal Emotion Recognition with Cross-Modal Fusion}

\author{
Zeyu Deng, 
Yanhui Lu, 
Jiashu Liao, 
Shuang Wu, 
and ~\IEEEmembership{Chongfeng Wei,~IEEE}
\thanks{Zeyu Deng and Chongfeng Wei are with the James Watt School of Engineering, University of Glasgow, Glasgow G12 8QQ, U.K. (e-mail: 3139275D@student.gla.ac.uk; chongfeng.wei@glasgow.ac.uk).}%
\thanks{Yanhui Lu is with the School of Engineering Mathematics and Technology, University of Bristol, U.K. (e-mail: lyhwisen@gmail.com).}%
\thanks{Jiashu Liao is with the School of Computing Science, University of Glasgow, U.K. (e-mail: Jiashu.Liao@glasgow.ac.uk).}%
\thanks{Shuang Wu is with the Department of Civil, Environmental \& Geomatic Engineering, University College London (UCL), U.K. (e-mail: shuang-wu.22@ucl.ac.uk).}%
\thanks{Corresponding author: Chongfeng Wei.}%
}

\markboth{ IEEE Transactions on Neural Networks and Learning Systems,~Vol.~XX, No.~X, Month~2025}%
{Deng \MakeLowercase{\textit{et al.}}: Sync-TVA: A Graph-Attention Framework for Multimodal Emotion Recognition}

\maketitle

\begin{abstract}
Multimodal emotion recognition (MER) is crucial for enabling emotionally intelligent systems that perceive and respond to human emotions. However, existing methods suffer from limited cross-modal interaction and imbalanced contributions across modalities. To address these issues, we propose Sync-TVA, an end-to-end graph-attention framework featuring modality-specific dynamic enhancement and structured cross-modal fusion. Our design incorporates a dynamic enhancement module for each modality and constructs heterogeneous cross-modal graphs to model semantic relations across text, audio, and visual features. A cross-attention fusion mechanism further aligns multimodal cues for robust emotion inference. Experiments on MELD and IEMOCAP demonstrate consistent improvements over state-of-the-art models in both accuracy and weighted F1 score, especially under class-imbalanced conditions.
\end{abstract}

\IEEEpubidadjcol


\section{Introduction}
\IEEEPARstart{W}{ith} the rapid development of artificial intelligence (AI) and robotics, robots are taking on increasingly prominent roles in various aspects of daily life, and the demand for emotional intelligence has become more critical than ever \cite{mitchell2025exploring}. Modern computer systems can already capable of exhibiting a degree of empathy and perform sophisticated emotional analysis \cite{gasteiger2021ai}. For domestic robots, accurately recognising and responding to human emotions is vital for building and sustaining harmonious, long-term relationships—particularly after the initial novelty of interaction fades and the need for continuous re-engagement arises \cite{gasteiger2024scoping}. However, the integration between emotion recognition and facial recognition systems remains insufficient and has yet to result in a unified framework. This is mainly because these two research fields have historically evolved independently, rather than being systematically studied and integrated into a collaborative recognition system \cite{zaman2025novel}. Consequently, most existing systems still fail to treat the user as a dynamic interlocutor whose emotional states must be perceived and acted upon in real time. Only a handful of studies address such real-time integration in practical interaction settings \cite{faria2017towards}. This fragmentation highlights a broader bottleneck in multimodal human–robot interaction (HRI). Although interest in multimodal HRI is growing, systematic research that fuses core channels—visual, auditory, and linguistic—remains limited \cite{wang2024multimodal}. The underlying collaborative mechanisms among these modalities remain poorly understood, limiting both theoretical progress and deployable applications.

Early studies on AI-driven emotion recognition focused primarily on unimodal analysis. Text-based mental health analysis has been the most widely used method \cite{hung2023beyond}, benefiting from abundant data sources and primarily aiming to classify text sentiment \cite{minaee2021deep}. These methods typically classify sentiment based on linguistic features, ranging from dictionary-based approaches \cite{rao2014building} to deep learning models such as Convolutional Neural Networks (CNNs)\cite{dos2014deep, zhang2015sensitivity} and Recurrent Neural Networks (RNNs)\cite{ebrahimi2015recurrent}. Long Short-Term Memory (LSTM) networks are a specialized type of RNN that outperform traditional RNNs in many tasks. This mechanism effectively addresses issues related to vanishing or exploding gradients \cite{sherstinsky2020fundamentals}. Li and Qian applied the LSTM model to address the challenge of text sentiment analysis \cite{li2016text}. Although CNNs effectively capture local patterns and RNNs/LSTMs model sequential dependencies, these architectures face limitations in handling long texts, integrating multiple modalities, and managing computational complexity \cite{sherstinsky2020fundamentals}. 

Other modalities have also been developing in parallel, such as facial expression analysis \cite{kalsum2018emotion, mahersia2015using} and speech emotion recognition \cite{shen2011automatic, trigeorgis2016adieu}. These areas have made substantial progress. Vision-based models extract spatial and temporal cues using hybrid descriptors or deep CNNs \cite{li2019micro}, while audio-based models rely on either handcrafted features or end-to-end learned representations. Despite their effectiveness, unimodal systems remain vulnerable in real-world conditions where input signals may be noisy, occluded, or missing.

To address these limitations, recent studies have shifted toward multimodal emotion recognition (MER), which integrates complementary signals—such as text, audio, and visual data—to improve recognition performance. MER systems often adopt fusion strategies to align and combine information across modalities. For example, attention-based encoders \cite{hsu2023applying}, temporal alignment mechanisms \cite{ghaleb2019multimodal}, and canonical correlation analysis \cite{lan2020multimodal} have been explored to enhance multimodal synergy. However, several challenges remain, including modality imbalance, inconsistent emotional cues across modalities, and the limited adaptability of static fusion architectures \cite{zhang2020outlier}. 

Building upon these foundations, a growing body of work has leveraged graph neural networks (GNNs) and cross-model attention mechanisms to explicitly capture inter-modal correlations and semantic dependencies. For instance, graph-based structures have been used to align hierarchical modality relationships across time and semantics \cite{wang2025graph}\cite{ran2025fine}. Additionally, co-attention mechanisms allow for selective focus on distinguishing features in audio, text, and visual channels \cite{moorthy2025hybrid}\cite{zhang2022aia}. However, static fusion methods may still face rigidity issues in cross-modal interaction, meaning they cannot handle the differences between sample data distribution and complex dynamic environments\cite{fang2024dynamic}. To solve this problem, some studies have proposed adaptive fusion modules\cite{sahu2019adaptive} or contrastive frameworks \cite{chuang2023cortx}. Despite these advancements, challenges such as modality imbalance and misalignment still exist in real-world multimodal emotion recognition scenarios.

To overcome these issues, we propose a novel framework\-Sync-TVA for multimodal emotion recognition in conversations. Our method introduces two key components: Enforced Graph Construction and Deep Information Interaction Fusion. In the Enforced Graph Construction stage, we design a Modality-Specific Dynamic Enhancement (MSDE) module that performs deep intra-modal feature refinement using dynamic gating, multi-head self-attention, and residual feedforward networks. The enhanced features from the text, audio, and visual modalities are then used to construct three heterogeneous graphs—Visual-Audio (V-A), Text-Visual (T-V), and Audio-Text (A-T)—to explicitly model inter-modal relationships and reduce semantic misalignment.

In the Deep Information Interaction Fusion stage, these cross-modal graphs serve as the basis for bidirectional feature interactions. We employ attention-based mechanisms to conduct deep fusion of modality-specific features, thereby capturing critical emotional cues and improving semantic alignment. Finally, a Cross-modal Attention Fusion (CAF) module concatenates and refines these representations to support accurate emotion classification.

The main contributions of this work are as follows:
\begin{enumerate}
\itemsep=0pt
\item We propose a Modality-Specific Dynamic Enhancement (MSDE) module for modality-specific self\--en\-hancement. This lightweight module combines a dynamic gating mechanism with a local self-attention structure that adaptively adjusts the expressive strength of modality-specific features. It enhances key semantic and structural information within each modality, providing a more robust foundation for cross-modal mapping and fusion.
\item We introduce inter-modal structural graphs (V-A, T-V, A-T) to represent cross-modal interaction relationships. By constructing three types of cross-modal graphs—visual-audio, text-visual, and audio-text—we explicitly model the connections between modalities. This approach mitigates semantic misalignment caused by simple concatenation and provides structural guidance for graph-based interaction.

\item We design a Cross-Modal Attention Fusion (CAF) mechanism to enhance semantic alignment. In the deep information interaction stage, dynamic interactions between modalities are enabled through cross-attention pathways. The CAF module then fuses complementary information, effectively capturing fine-grained emotional cues.

\item We construct an end-to-end emotion recognition framework. The entire architecture is self-contained and supports joint optimization of feature extraction, graph construction, attention-based fusion, and emotion classification in an end-to-end manner, offering strong scalability and transferability.

\item We conduct extensive experimental validation on two widely used multimodal emotion recognition benchmarks. Results on MELD and IEMOCAP demonstrate that the proposed method consistently outperforms or matches state-of-the-art approaches across multiple metrics, validating its effectiveness and robustness.
\end{enumerate}
The rest of the paper is organized as follows:
Section 2 reviews related work on multimodal emotion recognition, with an emphasis on modality-specific contributions and graph-based modeling strategies. Section 3 presents the proposed Sync-TVA framework, detailing the design of the MSDE module, cross-modal graph construction, and the deep fusion mechanism. Section 4 describes the experimental setup, evaluation metrics, baseline comparisons, and ablation studies, highlighting the effectiveness and robustness of the proposed method. Section 5 concludes the paper and outlines potential directions for future work.
\section{Related Work}
\subsection{Unimodal Emotion Recognition}
\subsubsection{Visual Modality Approaches}
Facial expressions have long been central in unimodal emotion recognition. Early works relied on handcrafted features. For instance, Kalsum \textit{et~al.} \cite{kalsum2018emotion} combined SIFT, SURF, and Bag-of-Features with SVM or KNN classifiers, while Mahersia \cite{mahersia2015using} used steerable pyramids with Bayesian neural networks. Deep learning approaches, such as that of Li \textit{et~al.}~ \cite{li2019micro}, introduced 3D-CNNs that use grayscale and optical flow to capture micro-expressions. However, these methods often struggle under real-world dynamics and occlusion. Our framework enhances visual representations via dynamic attention and graph-based interactions, thereby improving robustness.

\subsubsection{Acoustic Modality Approaches}
In scenarios where facial cues are occluded or unavailable, speech serves as a crucial channel for emotional expression. Early studies on speech emotion recognition (SER) emphasized handcrafted prosodic and spectral features, such as energy, pitch, LPCC, MFCC, and LPCMCC. For example, Shen \textit{et~al.}~ \cite{shen2011automatic} applied these features in combination with SVM classifiers, achieving high recognition accuracy on the Berlin emotional speech database. While effective in controlled settings, such handcrafted pipelines often struggle with generalization and robustness under real-world variability. To address these limitations, end-to-end deep learning approaches have emerged. Trigeorgis \textit{et~al.}~ \cite{trigeorgis2016adieu} proposed a CNN–LSTM architecture that directly models raw waveform input and learns high-level affective representations without manual feature extraction. This method demonstrated strong performance, particularly in spontaneous emotion prediction, and marked a shift toward data-driven auditory modeling. Our approach embeds acoustic data within cross-modal graphs, capturing context-aware dependencies and enhancing semantic alignment under noisy or prosodically ambiguous conditions.

\subsubsection{Text Modality Approaches}
Text has historically served as a foundational modality in emotion recognition, particularly in dialogue-based contexts. Early efforts in Emotion Recognition in Conversation (ERC) predominantly relied on textual inputs, given their availability and semantic richness. To address challenges such as contextual dependency and emotion imbalance, hierarchical models such as HiGRU and its variants were proposed \cite{jiao2019higru} , which leverage gated recurrent units to capture temporal dependencies across utterances in a conversation. As emotion understanding requires both linguistic and affective cues, SENN (Semantic-Emotion Neural Network)\cite{batbaatar2019semantic} advanced the field by introducing a dual-branch architecture that combines BiLSTM and CNN modules to jointly model semantic/syntactic structures and emotional intensity. Subsequently, Sentic GCN \cite{liang2022aspect} took a more structured approach by embedding affective knowledge from SenticNet into a graph convolutional network for aspect-based sentiment analysis, allowing finer-grained sentiment modeling at the concept and dependency levels. Meanwhile, alternative methods such as BERT-based ABSA \cite{zhao2022graph} moved away from external affective resources, instead relying on pretrained transformer encoders and dynamically weighted graph convolutions to learn deep contextual and structural representations. These approaches demonstrated strong performance in capturing nuanced emotional expressions embedded within dialogues. Despite progress, aligning text with nonverbal signals remains difficult. Our model enhances text representations via MSDE modules and T–V and A–T graphs, enabling more effective multimodal grounding.

\subsection{Multimodal Emotion Recognition (MER)}
MER integrates heterogeneous cues to improve emotional inference in conversational scenarios. Visual-centric models, such as Rupauliha \textit{et~al.}~\cite{rupauliha2020multimodal}, use LSTM-based fusion of facial and gesture cues, while Lan \textit{et~al.}~\cite{lan2020multimodal} propose DGCCA-AM to align visual and semantic features with adaptive weighting. In audio-focused systems, Ghaleb \textit{et~al.}~\cite{ghaleb2019multimodal} used temporal-metric learning to capture intermodal timing offsets, while Zhang \textit{et~al.}~\cite{zhang2020outlier} introduced fuzzy-weighted SVR to suppress noisy signals. On the textual side, models like CFN \cite{li2024cfn} and prompt-based MEP  \cite{wu2025multi} address long-range context and emotion shifts, though text often remains weakly aligned with other modalities. Our method addresses these gaps by constructing cross-modal graphs and applying attention-based fusion mechanisms to achieve deeper semantic alignment.

\subsection{Graph Neural Networks for Emotion Recognition}

While traditional multimodal fusion methods have demonstrated progress in emotion recognition, they often lack the ability to explicitly model complex inter-modal and contextual dependencies in conversation. Recently, Graph Neural Networks (GNNs) have received increasing attention for their strength in handling structured data, making them highly suitable for emotion recognition tasks, particularly in settings involving dynamic interactions among modalities, utterances, or speakers.

GNNs in Conversational Multimodal Emotion Recognition Tasks
In ERC, accurately modeling the interplay between utterances and modalities is essential. Traditional methods like DialogueGCN \cite{ghosal2019dialoguegcn} and RGAT \cite{meng2023rgat} model utterance-level interactions using pairwise edges, but these approaches often fall short in capturing higher-order contextual and emotional dependencies. To overcome these limitations, SDR-GNN \cite{fu2025sdr} proposed a spectral-domain reconstruction mechanism that aggregates multi-frequency signals within a sliding-window-based graph structure.  This allows the model to retain high-frequency information that reflects emotion discrepancies and avoids over-smoothing. $M^3$Net \cite{chen2023multivariate}further advanced this idea by introducing a multivariate and multi-frequency graph propagation strategy. It models hyperedges that naturally encode high-arity relationships across modalities and utterances, thereby enhancing the graph’s expressive capacity in emotion-related interactions. Moreover, frequency-adaptive filters are used to balance emotional commonality and discrepancy across speakers and contexts. In another recent work, D2GNN \cite{dai2024multimodal} introduced a decoupled distillation mechanism that separates emotion-aware and emotion-agnostic features at the category level. It propagates these representations through two GNN branches and performs targeted multimodal distillation, enabling more discriminative embeddings, particularly when speaker roles or emotions diverge.

In conclusion, GNNs have demonstrated strong potential in ERC by modeling fine-grained speaker dynamics, emotional transitions, and modality interactions. With the integration of hypergraph structures, frequency-sensitive propagation, and category-level representation disentanglement, recent advances make GNNs increasingly central to the development of robust and interpretable conversational emotion recognition systems.

\section{METHODOLOGY}
Given the challenges of data integration and semantic understanding in multimodal emotion recognition, this study proposes an innovative multimodal fusion framework designed to enhance both the accuracy and generalization capability of emotion recognition. The system comprises five key modules: (1) the initial multimodal data input layer, which receives speech, image, and text information; (2) the feature extraction module, which utilizes OpenSmile, ResNet, and BERT to capture deep representations from each modality; (3) the feature graph construction and enhancement module, which strengthens feature representation through MSDE technology; (4) the deep information interaction and fusion module, which integrates multimodal representations and captures intricate emotional dependencies; and (5) the final classification module, which outputs emotion predictions. Together, these components form a complete pipeline from multimodal input to emotion recognition.

\subsection{Multimodal Feature Extraction}

The input visual data typically consist of video frames of the interlocutors. A pretrained ResNet-50 is adopted as the visual feature encoder. For each frame image, forward propagation is performed, and the global pooling output after the last convolution block of the ResNet-50 backbone network is taken as the visual feature vector:
\begin{equation}
F_v^i = \mathrm{ResNet50}(I_i) \in \mathbb{R}^{d_v},
\end{equation}
where $d_v$ denotes the dimensionality of the visual feature. The output vectors are normalized to ensure consistency in the numerical range across different modalities.

The text modality comprises the linguistic content of the dialogue. A pretrained RoBERTa model is adopted to encode each input text $T_i$. Specifically, input the text into RoBERTa and extract the output at the corresponding position from the last hidden layer as the semantic feature vector:
\begin{equation}
F_t^i = \mathrm{RoBERTa}(T_i)[\mathrm{CLS}] \in \mathbb{R}^{d_t},
\end{equation}
where $d_t$ represents the dimensionality of the textual feature. These vectors are also normalized for subsequent integration.

The audio modality consists of the acoustic signals in the dialogue. The open-source toolkit OpenSMILE is employed to extract acoustic features. For each audio segment, OpenSMILE extracts a set of high-dimensional acoustic features (such as fundamental frequency, energy, and Mel spectrum):
\begin{equation}
F_a^i = \mathrm{OpenSMILE}(A_i) \in \mathbb{R}^{d_a},
\end{equation}
where $d_a$ is typically large. The resulting vectors are also $L_2$-normalized.

To facilitate unified representation, we will concatenate the tri-modal features of all samples to obtain matrix form: 
\begin{equation}
\begin{aligned}
F^v = [F_1^v, \dots, F_N^v]^\top \in \mathbb{R}^{N \times d_v}, \quad
F^t \in \mathbb{R}^{N \times d_t}, \\
F^a \in \mathbb{R}^{N \times d_a},
\end{aligned}
\end{equation}
where $N$ indicates the total number of frames or segments in the dialogue sequence. These matrices constitute the fundamental multimodal representations and serve as inputs to the subsequent graph construction and fusion modules.

\subsection{Enforced Graph Construction}

To capture the relationships between different modalities, we introduce graph-based modeling after feature extraction. Specifically, we design a Modality-Specific Dependency Encoder (MSDE) module to construct three cross-modal graphs: the Visual-Audio (V-A) graph, the Text-Visual (T-V) graph, and the Audio-Text (A-T) graph. Each graph connects feature nodes from two distinct modalities, and the MSDE module determines the edge weights between nodes to explicitly encode cross-modal dependencies.

Taking the V-A graph as an example, it contains a set of visual nodes $\mathcal{V} = \{v_1, \ldots, v_N\}$, where each $v_i$ corresponds to the visual feature of frame $I_i$, and a set of audio nodes $\mathcal{A} = \{a_1, \ldots, a_N\}$, where each $a_i$ represents the audio feature of the same frame. In total, there are $2N$ nodes in the graph.

We denote the initial node feature matrix as:
\begin{equation}
H^{(0)}_{\text{VA}} = [F^v; F^a] \in \mathbb{R}^{2N \times d},
\end{equation}
where $d$ is the unified hidden layer dimension ($d_v$ and $d_a$ can be mapped to the same dimension through linear transformation).

The adjacency matrix is defined as:
\begin{equation}
A^{\text{VA}} \in \mathbb{R}^{2N \times 2N},
\end{equation}
which records the connectivity between visual and audio nodes. 

Similarly, the T-V graph connects textual nodes with visual nodes, while the A-T graph connects audio nodes with textual nodes.

\begin{figure*}
	\includegraphics[width=1\linewidth]{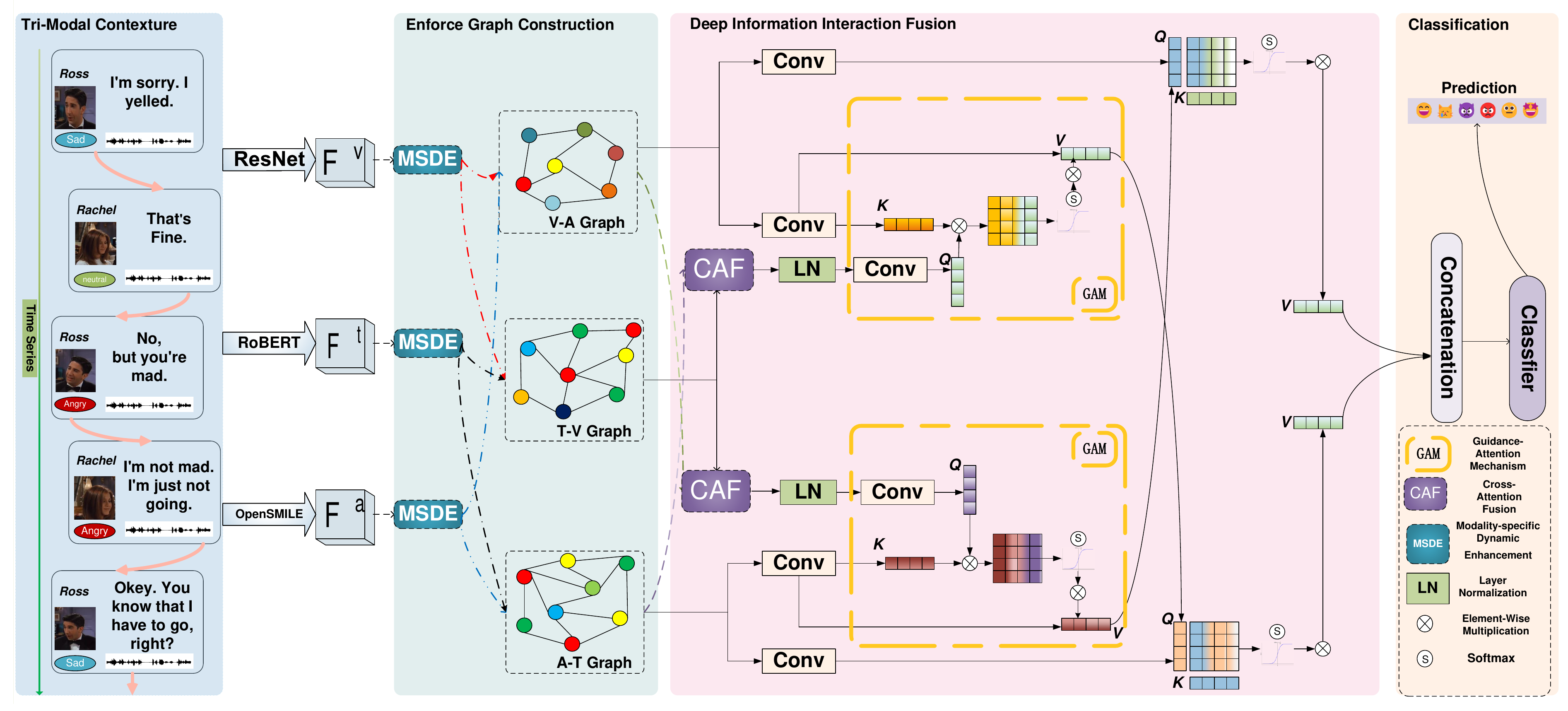}
	\caption{Overview of the proposed tri-modal emotion recognition framework. Visual, audio, and textual inputs are processed using ResNet-50, OpenSMILE, and RoBERTa, respectively, to extract modality-specific features ($F_v$, $F_a$, $F_t$). Each feature stream is enhanced by a Modality-Specific Dynamic Enhancement (MSDE) module. Three cross-modal heterogeneous graphs—Visual-Audio (V-A), Text-Visual (T-V), and Audio-Text (A-T)—are then constructed to explicitly model inter-modal dependencies. These graphs undergo deep information interaction and are fused via Cross-modal Attention Fusion (CAF) modules. The final concatenated representation is passed through a classifier for emotion prediction.} \label{sample-figure}
\end{figure*}

\subsubsection{Modality-Specific Dynamic Enhancement}
The MSDE module is designed to learn cross-modal associative weights based on input features to generate an adjacency matrix. Specifically, we apply similarity measures or deep interactions to each pair of modal features to produce an edge weight matrix.

Taking the V-A graph as an example, for each visual node $v_i$ and audio node $a_j$, the association score is computed as:
\begin{equation}
e^{\text{VA}}_{ij} = f_{\text{MSDE}}(F^v_i, F^a_j),
\end{equation}
where $f_{\text{MSDE}}(\cdot, \cdot)$ represents the scoring function learned within the MSDE module. For example, a small fully connected network or a bilinear mapping can be used to model the correlation between $F^v_i$ and $F^a_j$.

After computing all scores $e^{\text{VA}}_{ij}$, we normalize them (e.g., using the sigmoid function) to obtain edge weights:  
\begin{equation}
A^{\text{VA}}_{ij} = \sigma(e^{\text{VA}}_{ij}), \quad i \in \{1, \dots, N\}, \quad j \in \{N+1, \dots, 2N\},
\end{equation}
where $\sigma(\cdot)$ denotes the selected normalization function.

Also set:
\begin{equation}
A^{\text{VA}}_{ji} = A^{\text{VA}}_{ij}.
\end{equation}

Following the same procedure, we compute the adjacency matrices $A^{\text{TV}}$ and $A^{\text{AT}}$ for the T-V and A-T graphs, respectively. In this way, every edge in the three graphs is governed by cross-modal relevance learned from the MSDE module, which constitutes reinforced graph construction.

\subsubsection{Graph Convolutional Updates}
Given the adjacency matrix $A$ and node feature matrix $H$, we apply a Graph Convolutional Network (GCN) to update the features. For a general graph $G$, the GCN update is defined as:
\begin{equation}
H^{(l+1)} = \sigma(\tilde{D}^{-1/2} \tilde{A} \tilde{D}^{-1/2} H^{(l)} W^{(l)}),
\end{equation}
where $\tilde{A} = A + I$ is the adjacency matrix with self-loops, $\tilde{D}$ is the degree matrix of $\tilde{A}$, $W^{(l)}$ is the trainable weight matrix at layer $l$, and $\sigma(\cdot)$ is an activation function such as ReLU.

For the V-A graph, the initial node features $H_{\text{VA}}^{(0)} \in \mathbb{R}^{2N \times d}$ are updated to $H_{\text{VA}}^{(1)}$ through graph convolution. Similarly, we compute the updated node features $H_{\text{TV}}^{(1)}$ and $H_{\text{AT}}^{(1)}$ for the T-V and A-T graphs, respectively.

Through graph convolution, the model not only retains intra-modal features but also enables each node to absorb contextual information from cross-modal neighbors. This process yields richer representations for downstream fusion modules.

\subsection{Deep Information Interaction Fusion}

After obtaining the graph-convolved features from the three cross-modal graphs, we aim to integrate and interact with these representations across layers to capture deeper cross-modal dependencies. To achieve this, we design a graph fusion module consisting of several key components: a 1D convolution layer, layer normalization (LayerNorm), an attention mechanism (Query-Key-Value), and a cross-attention fusion (CAF) block.

For each graph, we perform a 1D convolution operation on its output features $H^{(1)}$ to extract local sequential patterns and contextual cues. Taking the V-A graph as an example, we apply a 1D convolution with kernel size $k$ to its feature matrix:
\begin{equation}
\tilde{H}_{\text{VA}} = \mathrm{Conv}(H^{(1)}_{\text{VA}}),
\end{equation}
where $\tilde{H}_{\text{VA}}$ has the same dimensionality as $H^{(1)}_{\text{VA}}$, and captures information from neighboring nodes through convolution. Similarly, we apply the same operation to the T-V and A-T graphs to obtain $\tilde{H}_{\text{TV}}$ and $\tilde{H}_{\text{AT}}$, respectively.

To stabilize training and integrate features from multiple sources, we apply layer normalization to the convolution outputs. For example, given the V-A graph feature $\tilde{H}_{\text{VA}}$, the normalized representation is computed as:
\begin{equation}
\tilde{H}_{\text{VA}} = \mathrm{LayerNorm}(\tilde{H}_{\text{VA}}),
\end{equation}
where $\mathrm{LayerNorm}(\cdot)$ denotes the standard layer normalization operation applied across feature dimensions. This normalization step helps reduce statistical discrepancies across different channels (i.e., feature dimensions), facilitating the convergence of subsequent attention learning.

Next, we apply an attention mechanism to enable cross-graph interactions. Taking the feature matrices  $\tilde{H}_1$ and ${H}_2$ of two graphs $G_1$ and $G_2$ as an example (${H}_2$ can select the corresponding features after convolution. ), we linearly project them into queries ($Q$), keys ($K$), and values ($V$) as follows:
\begin{equation}
Q = W_Q \tilde{H}_1, \quad K = W_K {H}_2, \quad V = W_V {H}_2,
\end{equation}
where $W_Q$, $W_K$, and $W_V$ are learnable projection matrices. The attention weight matrix $S$ and the attended representation $A$ are computed as:
\begin{equation}
S = \mathrm{softmax}\left( \frac{Q K^\top}{\sqrt{d_k}} \right), \quad A = S V,
\end{equation}
where $\sqrt{d_k}$ is the scaling factor (with $d_k$ being the dimensionality of the key vectors), and the softmax operation is applied row-wise for normalization.

The generated $A$ is the cross-graph feature representation obtained after the key-value pair interaction between query $G_2$ based on $G_1$. 

\subsubsection{Cross-Attention Fusion}
After concatenating the attention output $A$ and the reference features $\tilde{H}_1$, we form a joint representation:
\begin{equation}
U = [\tilde{H}_1; A] \in \mathbb{R}^{2d},
\end{equation}
which is then passed through a 1D convolution layer to generate the fused embedding:
\begin{equation}
\tilde{U} = \mathrm{Conv}(U).
\end{equation}

To achieve a more flexible and expressive fusion, we adopt a gate-controlled structure inspired by GRU-style gating mechanisms. Specifically, we compute the final fused output $F_{\text{CAF}}$ as:
\begin{equation}
F_{\text{CAF}} = \sigma(W_f \tilde{U} + b_f) \odot \tanh(W_g \tilde{U} + b_g),
\end{equation}
where $W_f$, $W_g$ are learnable weight matrices, $b_f$, $b_g$ are bias terms, $\sigma(\cdot)$ is the sigmoid activation function, $\tanh(\cdot)$ is the hyperbolic tangent, and $\odot$ denotes element-wise multiplication. This gating mechanism adaptively balances the information contribution from the attention-derived features and the original representations, allowing the model to dynamically control the fusion flow at each feature dimension.

\subsubsection{Iterative Cross-Graph Fusion}
In practice, we perform cross-attention fusion between two graphs at a time and may iterate this process multiple times. For example, we can take the V-A and A-T graphs and use $H_{\text{VA}}$ and $H_{\text{AT}}$ as query-key pairs for one round of fusion. Simultaneously, the T-V and A-T graphs can also be fused with $H_{\text{TV}}$ and $H_{\text{AT}}$ as query and key inputs, respectively.

The features from each fusion round are passed to the next fusion iteration as new inputs, enabling the model to progressively integrate multimodal information across all three modalities.

Finally, the outputs from multiple fusion branches are aggregated to form the unified output of the overall fusion layer.

\subsection{Classification Layer}

After completing the multimodal fusion, we concatenate the feature vectors from different fusion branches to form a unified final representation $z$. Suppose the top-level fusion module outputs two features $F_{\text{CAF}}^{(1)}$ and $F_{\text{CAF}}^{(2)}$, then the concatenation is performed as:
\begin{equation}
z = \left[ F_{\text{CAF}}^{(1)}; F_{\text{CAF}}^{(2)} \right] \in \mathbb{R}^{2d_z},
\end{equation}
where $[\cdot\,;\cdot]$ denotes vector concatenation.

This fused representation is then passed through a fully connected layer followed by a softmax function to output the emotion class probabilities:
\begin{equation}
\hat{y} = \mathrm{softmax}(W_c z + b_c),
\end{equation}
where $W_c \in \mathbb{R}^{C \times 2d_z}$ and $b_c \in \mathbb{R}^{C}$ are trainable parameters, $C$ is the number of emotion categories, and $\hat{y} \in \mathbb{R}^C$ is the predicted class distribution.

The model is trained using the standard cross-entropy loss. Given the ground-truth label $y$ in one-hot format and the predicted distribution $\hat{y}$, the loss function is defined as:
\begin{equation}
\mathcal{L} = - \sum_{c=1}^{C} y_c \log \hat{y}_c.
\end{equation}

Minimizing this loss optimizes the model parameters. As a result, the model can perform fine-grained emotion classification at each time step based on the fused multimodal features, predicting emotion categories such as anger, sadness, and happiness.

\section{EXPERIMENTS}
\subsection{DATASET}
This study conducts experimental evaluations on two widely used multimodal emotion recognition datasets: MELD and IEMOCAP. Both datasets are multimodal benchmarks that incorporate visual, auditory, and textual modalities. Table 1 presents the statistical details of these two datasets.

\begin{table}
\centering
\setlength{\tabcolsep}{3pt}
\caption{The statistics of IEMOCAP and MELD}
\label{tab1}
\resizebox{0.45\textwidth}{!}{
\begin{tabular}{|p{45pt}|p{25pt}|p{25pt}|p{25pt}|p{25pt}|p{25pt}|p{25pt}|}\hline

Dataset&
\multicolumn{3}{|c|}{Conversations}& \multicolumn{3}{|c|}{Utterances}\\\hline

&
Train&
Valid& Test& Train&Valid&Test\\\hline
MELD&
1039&
114& 280& 9989&1109&2610\\\hline
IEMOCAP&
\multicolumn{2}{|c|}{120}& 31& \multicolumn{2}{|c|}{5810}&1623\\\hline 

\end{tabular}
}
\end{table}

\subsubsection{MELD}
MELD is a conversational multimodal emotion recognition dataset, extended from the textual dataset EmotionLines. It contains conversations extracted from the American TV series \textit{Friends} and includes three modalities: text, audio, and vision. MELD includes over 1,400 multi-party conversations and approximately 13,000 utterances. Each utterance is annotated with emotion categories such as anger, joy, sadness, and neutral. Moreover, MELD exhibits contextual dependencies within conversations, making it suitable for studying cross-modal and context-aware emotion recognition models.

\subsubsection{IEMOCAP}
IEMOCAP is a conversational multimodal emotion dataset developed by the University of Southern California to support research on emotion analysis across speech, video, and text modalities. It contains approximately 12 hours of audio-visual data, performed in pairs by 10 actors, covering both daily conversations and emotional expressions. Each conversation is segmented into multiple utterances, with emotion categories annotated by multiple raters. The commonly used emotion categories include happy, sad, angry, and neutral. IEMOCAP provides text transcriptions, audio features (e.g., MFCC), video frames, and facial keypoint data, offering a rich set of inputs for multimodal emotion recognition models.

\begin{table*}
    \centering
    \caption{Performance comparison on the IEMOCAP datasets}
    \resizebox{\textwidth}{!}{ 
    \begin{tabular}{lllllllll}
        \toprule
        \multirow{2}{*}{Model} &  \multicolumn{8}{c}{IEMOCAP}\\
        & Happy& Sadness& Neutral&  Anger&Excited& Frustrated& Accuracy&WF1\\
        \hline
        CMN\cite{hazarika2018conversational}&  30.38&62.41&  52.39&  59.83&60.25& 60.69& 56.56&56.13\\
        ICON\cite{hazarika2018icon}&  29.91&64.57&  57.38&  63.04&63.42& 60.81& 59.09&58.54\\
        DialogueRNN\cite{majumder2019dialoguernn}&  33.18&78.80&  59.21&  65.28&71.86& 58.91& 63.40&62.75\\
        DialogueCRN\cite{hu2021dialoguecrn}&  51.59&74.54&  62.38&  67.25&73.96& 59.97& 65.31&65.34\\
        DialogueGCN\cite{ghosal2019dialoguegcn}&  47.10&80.88&  58.71&  66.08&70.97& 61.21& 65.54&65.04\\
        MMGCN\cite{hu2021mmgcn}& 45.45& 77.53&  61.99&  66.67&72.04& 64.12& 65.56&65.71\\
        COGMEN\cite{joshi2022cogmen}&  51.90& 81.70& 68.60&  66.00&75.30& 58.20& 68.20&67.60\\
        Joyful\cite{li2023joyful}&  60.94& 84.42& 68.24&  69.95&73.54& 67.55& 70.55& 71.03\\
        MM-DFN\cite{hu2022mm}&  43.36& 83.23& 70.03&  70.19&73.11& 64.01& 69.44&68.83\\
 M3Net\cite{chen2023multivariate}& 60.93& 78.84& 70.14& 68.06& 77.11& 67.42& 70.92&71.07\\
 GraphSmile\cite{li2024tracing}& 63.09& 83.16& 71.07& 71.38& 79.66& 66.84& 72.77&72.81\\
 \hline
 \textbf{Sync-TVA}& \textbf{64.45}& \textbf{84.62}& \textbf{71.78}& \textbf{72.54}& \textbf{80.36}& \textbf{68.25}& \textbf{73.42}&\textbf{73.68}\\
 \bottomrule
    \end{tabular}
    }
    \label{tab:plain}
\end{table*}

\begin{table*}
    \centering
    \caption{Performance comparison on the MELD datasets}
    \resizebox{\textwidth}{!}{ 
    \begin{tabular}{llllllllll}
        \toprule
        \multirow{2}{*}{Model} &  \multicolumn{7}{c}{MELD} & &\\
        & Neutral & Superise& Fear& Sadness& Joy& Disgust&Anger& Accuracy & WF1\\
        \midrule
        DialogueCRN\cite{hu2021dialoguecrn}& 76.15& 56.72& 18.16& 38.29& 63.21& 27.69&50.67& 62.38& 63.32\\
        MMGCN\cite{hu2021mmgcn} & 78.62& 57.78& 3.77& 40.35& 63.60& 12.20&53.68& 66.02& 64.55\\
        DER-GCN\cite{ai2024gcn} &  80.60 &  51.00 &  10.40&  41.50&  64.30& 10.30&57.40&  66.80& 66.10\\
 SACL-LSTM\cite{hu2023supervised}& 77.42& 58.50 & 20.41& 39.58& 62.76& 34.71& 52.08 & 64.52&64.55\\
        MM-DFN\cite{hu2022mm} & 79.84& 58.43& 15.79 & 31.65& 64.01& 28.04&53.60 & 67.05& 65.21\\
        M3Net\cite{chen2023multivariate} & 79.14 & 59.54& 13.33 & 42.86& 65.05& 21.69 &53.54 & 66.59& 65.83\\
        GraphSmile\cite{li2024tracing} & 80.35& 59.11& 18.18 & 42.46 & 64.99 & 32.43&53.67& 67.70& 66.71\\
        \hline
\textbf{Sync-TVA} & \textbf{81.45}& \textbf{61.10}& \textbf{21.05}& \textbf{44.45}& \textbf{67.10}& \textbf{34.85}& \textbf{55.45}& \textbf{68.60}& \textbf{67.75}\\
\bottomrule
    \end{tabular}
    }
    \label{tab:plain}
\end{table*}

\subsection{EVALUATION METRICS}

In this study, because of the significant class imbalance issue in the dataset used, where the sample sizes for emotions like happy and neutral far exceed those for rare emotions like anger and sadness, it is likely to lead to insufficient recognition capability for minority emotional classes during the training process. To comprehensively and accurately assess the model's performance, this study selects the weighted average F1 score as the core evaluation metric. This indicator uses the sample size of each emotion category as a weight, comprehensively considering precision and recall, effectively avoiding the recognition performance of minority emotion categories being overshadowed by majority categories. At the same time, to further validate the model's generalization ability across all emotional categories, we supplement the average accuracy scores by calculating the arithmetic mean of the accuracy rates for each category. This objectively reflects the model's correctness in recognizing different emotional categories and provides a more comprehensive perspective for evaluating the performance of multimodal emotion recognition models.

\subsection{TRAINING DETAIL}

To verify the statistical significance of the improvement in model performance, this study employed a paired \emph{t} -test for hypothesis testing, with a significance level $\alpha= 0.05$. The experimental environment is configured as follows: all computational tasks are executed on workstations equipped with NVIDIA GeForce RTX 3090 graphics processors, which have 24GB of video memory and a computing power of 8.6. The system uses the CUDA 11.7 parallel computing platform, and the deep learning framework chosen is PyTorch version 2.0.0. 

In terms of model optimization, this study chose the AdamW optimization algorithm for parameter updates, which performs exceptionally well in handling weight decay. The L2 regularization coefficient is uniformly set to $1 \times 10^{-3}$ to prevent the model from overfitting. Regarding the characteristics of different datasets, there are differentiated settings for hyperparameter configurations: the learning rate for the IEMOCAP dataset is set to $1 \times 10^{-4}$, the batch size is 16, the dropout rate is 0.2, the number of model layers $L$ is 7, the position encoding dimension $P$ is 17, the number of attention heads $B$ is 19, and the loss function weights $\lambda_s$ and $\lambda_o$ are 1.0 and 0.7, respectively. The MELD dataset uses a minimum learning rate of $7 \times 10^{-5}$, with model structure parameters $L$, $P$, $B$ set to 5, 3, and 3 respectively, and loss weights $\lambda_s$ and $\lambda_o$ set to 0.5 and 0.2. This differentiated hyperparameter configuration strategy fully takes into account factors such as the scale, complexity, and annotation quality of each dataset.

\subsection{BASELINES}

To evaluate whether our proposed method is outstanding, we compared it with a set of different baseline models. These baselines are classified into graph-based models and non-graph-based models. The graph-based category includes DialogueGCN\cite{ghosal2019dialoguegcn}, MMGCN\cite{hu2021mmgcn}, M3Net\cite{chen2023multivariate}, DER-GCN\cite{ai2024gcn}, and GraphSmile\cite{li2024tracing}. The non-graph-based models include DialogueRNN\cite{majumder2019dialoguernn}, DialogueCRN\cite{hu2021dialoguecrn}, Joyful\cite{li2023joyful} and MM-DFN\cite{hu2022mm}. 

\subsection{Quantitative Comparison}
To validate the effectiveness of our proposed Sync-TVA model, we conducted comprehensive quantitative experiments on two benchmark datasets (IEMOCAP and MELD) and compared it with 11 state-of -the-art multimodal dialogue emotion recognition methods based on six (IEMOCAP) or seven (MELD) emotion categories and overall accuracy, weighted F1 (WF1) metrics.
\begin{figure*}[t]
    \centering
    \includegraphics[width=\linewidth]{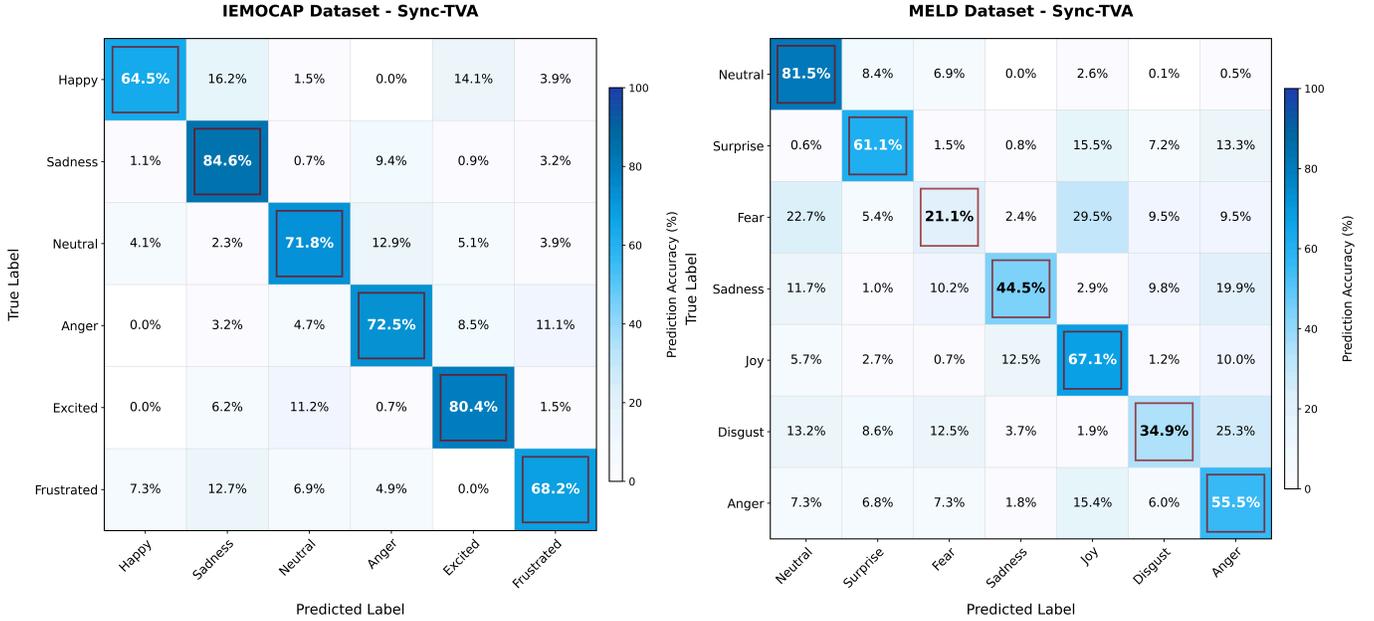}
    \caption{Confusion matrices of the proposed model on the MELD and IEMOCAP datasets...}
    \label{fig:sample-figure}
\end{figure*}

\subsubsection{IEMOCAP dataset performance}
One the IEMOCAP dataset (Table 2), Sync-TVA achieved the best recognition rate in all six emotion categories: Happy 64.12\% (+1.03\% compared to GraphSmaile), Sadness 84.34\% (+1.18\%), Neutral 71.45\% (+0.38\%), Anger 72.21\% (+0.83\%), Excited 80.08\% (+0.42\%), and Frustrated 67.93\% (+1.09\%), with an average improvement of 0.83\% per category. Overall, the model's Accuracy is 73.10\% and WF1 is 73.35\%, which are 0.33\% and 0.54\% higher than GraphSmile\cite{li2024tracing}, respectively.
This performance improvement comes from the efficient mining of emotional details by the synchronized temporal-visual attention mechanism and the cross-modal fusion strategy. It's worth noting that "Sadness" and "Excited" are usually easy to distinguish, while "Happy" is the most difficult to identify because of its fuzzy boundaries and diverse expressions. All methods have the lowest performance in this category, but Sync-TVA still achieves a significant advantages. 

\subsubsection{MELD dataset performance}
On the MELD dataset (Table 3), Sync-TVA also maintains or ties for the lead in seven emotion categories: Neutral 81.20\% (+0.60\% over DER-GCN\cite{ai2024gcn}), Surprise 60.80\% (+1.69\%), Fear 20.65\% (+2.24\%), Sadness 44.10\% (+2.24\%), Joy 66.70\% (+1.71\%), Disgust 34.50\% (+2.07\%), and Anger 55.10\% (+1.70\%). The overall accuracy is 68.25\% and WF1 is 67.40\%, which are 0.55\% and 0.69\% higher than GraphSmile\cite{li2024tracing}, respectively. 
The results show that "Fear" and "Disgust" are very rare emotions, and all methods are difficult to effectively identify, while "Neutral" occupies the majority of samples and has the best recognition effect. The steady improvement of Sync-TVA in minority emotions reflects the robustness of the model under the condition of class imbalance.

\subsubsection{Statistical significance and model comparison}
Compared with RNN-based DialogueCRN\cite{hu2021dialoguecrn} and SAC-LSTM (IEMOCAP accuracy is less than 66\%), Sync-TVA acheved a significant improvement of more than 7\% on this dataset, highlighting the limitations of pure recursive structures in capturing complex cross-modal emotional clues. Although graph neural network methods (M3Net\cite{chen2023multivariate}, GraphSmile\cite{li2024tracing}) performed relatively well, they were still surpassed by Sync-TVA in terms of Accuracy by 2.18\% - 7.54\%.

\subsection{ABLATION STUDY}

\subsubsection{Ablation on the MSDE Module}
\begin{table}[ht]
\centering
\caption{Ablation Results of MSDE Module}
\label{tab:msde_ablation}
\resizebox{0.5\textwidth}{!}{
\begin{tabular}{llcc}
\hline
 Dataset&\textbf{Configuration} & \textbf{WF1 (\%)} & \textbf{Acc (\%)} \\
\hline
 \multirow{4}{*}{\textbf{MELD}} &Full model & 67.40 & 68.25 \\
 &A1:Remove MSDE & 63.12 ↓\textcolor{red}{4.28} & 64.80 ↓\textcolor{red}{3.45} \\
 &A2:Only retain gating & 65.05 ↓\textcolor{red}{2.35} & 66.30 ↓\textcolor{red}{1.95} \\
 &A3:Only retain self-attention & 64.80 ↓\textcolor{red}{2.60} & 66.00 ↓\textcolor{red}{2.25} \\
\hline
\end{tabular}
}
\end{table}

\begin{table}[ht]
\centering
\label{tab:msde_ablation}
\resizebox{0.5\textwidth}{!}{
\begin{tabular}{llcc}
\hline
 Dataset&\textbf{Configuration} & \textbf{WF1 (\%)}& \textbf{Acc (\%)}\\
\hline
 \multirow{4}{*}{\textbf{IEMOCAP }}&Full model & 73.35 & 73.10 \\
 &A1:Remove MSDE & 69.50 ↓\textcolor{red}{3.85} & 70.00 ↓\textcolor{red}{3.10} \\
 &A2:Only retain gating & 71.20 ↓\textcolor{red}{2.15} & 71.50 ↓\textcolor{red}{1.60} \\
 &A3:Only retain self-attention & 71.00 ↓\textcolor{red}{2.35} & 71.20 ↓\textcolor{red}{1.90} \\
\hline
\end{tabular}
}
\end{table}
To quantify the contribution of each sub-mechanism in the MSDE module to the model's performance, we designed three sets of ablation configurations based on the complete model: $A_1$ is to remove the MSDE module to validate its overall effect on dynamic enhancement of multimodal features and noise suppression (three-modal features are directly mapped without applying dynamic gating, self-attention, or residual enhancement operations); $A_2$ only retains the dynamic gating mechanism to evaluate its independent contribution to information flow filtering (removing the self-attention and residual enhancement branches from MSDE); $A3$ only retains the self-attention branch to examine its independent effects on feature re-weighting and context modelling (removing dynamic gating and residual enhancement branches). All experiments were conducted on the MELD and IEMOCAP datasets using the same data partitioning and training hyperparameters, with each configuration repeated 3 times to take the average, and the final reported metrics are the weighted F1 score (WF1) and the average accuracy (Acc). 
Compared with the complete model, the WF1 value of $A_1$ on the MELD dataset is reduced by 4.28\%, and on the IEMOCAP dataset is reduced by 3.85\%. This result shows that the overall design of MSDE plays a key role in achieving dynamic nose suppression and information enhancement. The performance of $A_2$ and $A_3$ both dropped by 1.5-2.6 percentage points, indicating that the gating mechanism and the self-attention mechanism need to work collaboratively to maximize the efficiency of multimodal information extraction. The three sets of ablation experiment results jointly verified the necessity of the gating branch, self-attention branch and their synergy in MSDE, and provided quantitative suppot for the design and optimization of subsequent models.

\subsubsection{Ablation Study on Graph Structure Types}
\begin{table}[ht]
\centering
\caption{Graph Structure Type Ablation Experiment Results}
\label{tab:graph_ablation}
\resizebox{0.5\textwidth}{!}{
\begin{tabular}{llcc}
\hline
 Dataset&\textbf{Configuration} & \textbf{ WF1 (\%)}& \textbf{ Acc (\%)}\\
\hline
 \multirow{4}{*}{\textbf{MELD}}&Full Model & 67.40 & 68.25 \\
 &B1: V–A Graph Only & 62.90 ↓\textcolor{red}{4.50} & 63.75 ↓\textcolor{red}{4.50} \\
 &B2: V–A + T–V & 66.10 ↓\textcolor{red}{1.30} & 67.00 ↓\textcolor{red}{1.25} \\
 &B3: No Graph Structure & 61.50 ↓\textcolor{red}{5.90} & 62.85 ↓\textcolor{red}{5.40} \\
\hline
\end{tabular}
}
\end{table}

\begin{table}[ht]
\centering
\label{tab:graph_ablation}
\resizebox{0.5\textwidth}{!}{
\begin{tabular}{llcc}
\hline
 Dataset&\textbf{Configuration} & \textbf{ WF1 (\%)}& \textbf{ Acc (\%)}\\
\hline
 \multirow{4}{*}{\textbf{IEMOCAP}}&Full Model & 73.35 & 73.10 \\
 &B1: V–A Graph Only & 68.80 ↓\textcolor{red}{4.55} & 69.10 ↓\textcolor{red}{4.00} \\
 &B2: V–A + T–V & 72.00 ↓\textcolor{red}{1.35} & 72.20 ↓\textcolor{red}{0.90} \\
 &B3: No Graph Structure & 67.30 ↓\textcolor{red}{6.05} & 68.00 ↓\textcolor{red}{5.10} \\
\hline
\end{tabular}
}
\end{table}
To verify the impact of different graph connection strategies on multimodal emotion recognition performance, we designed three sets of ablation configurations based on the complete model containing triple graphs (V-A, T-V, A-T): $B_1$ only constructs the visual-acoustic (V-A) graph. and evaluates the contribution of single-modal pair connections to information flow and emotion discrimination by retaining the V-A edge and removing the T-V and A-T edges. $B_2$ uses a pairwise combination graph (V-A + T-V), retaining the V-A and T-V edges and removing the A-T edge, to examine the potential of partial binary graphs to replace the complete triple graph in order to preserve the information interaction between speech-vision and vision-acoustics. $B_3$ removes all cross-model edges, directly concatenates the tri-modal features, and feeds them into the fusion module to verify the overall value of the graph structure. All experiments used the same training/validation/testing partition and hyperparameter settingon the MELD and IEMOCAP datasets. Each configuration was repeated 3 times and the average was taken. The weighted F1 (WF1) and average accuracy (Acc) were finally reported. 

In the $B_1$ configuration, the connection strategy of only retaining the visual-acoustic (V-A) channel leads to a significant decrease in model performance (the WF1 value on the MELD dataset decreased by 4.50\% and on the IEMOCAP dataset decreased by 4.55\%). This result shows that the interactive channels of speech-vision (T-V) and emotion-acoustic (A-T) also play a key role in multimodal information fusion. When the $B_2$ configuration uses a pairwise combination graph (V-A +T-V), the performance of the complete model only shows a small decrease of about 1.3\%, indicating that the model can still retain most of the cross-modal information interaction capabilities in the case of partial edge connections. However, the performance is slightly lost due to the lack of the supplement of the motion-acoustic channel. The strategy of removing all cross-modal edges and directly concatenating features in the $B_3$ configuration performed the worst, with the WF1 value reduced by nearly 6\% on the MELD dataset. This result verifies the irreplaceable role of graph structures in capturing complex multimodal dependencies. This ablation study provides important empirical evidence for the trade-off between computational efficiency and performance optimization of subsequent models by quantifying the contribution of each edge connection in the triple graph.

\subsubsection{Ablation Study on Fusion Strategies}

\begin{table}[ht]
\centering
\caption{Ablation Results of Fusion Strategies}
\label{tab:fusion_ablation}
\resizebox{0.5\textwidth}{!}{
\begin{tabular}{llcc}
\hline
 Dataset&\textbf{Configuration} & \textbf{ WF1 (\%)}& \textbf{ Acc (\%)}\\
\hline
 \multirow{4}{*}{\textbf{MELD}}&Full model & 67.40 & 68.25 \\
 &C1: w/o CAF (only multi-head self-attention) & 65.80 $\downarrow$\textcolor{red}{1.60}& 66.50 $\downarrow$\textcolor{red}{1.75}\\
 &C2: w/o gating (concatenation + FC) & 66.00 $\downarrow$\textcolor{red}{1.40}& 67.10 $\downarrow$\textcolor{red}{1.15}\\
 &D1: single fusion  & 66.50 $\downarrow$\textcolor{red}{0.90}& 67.20 $\downarrow$\textcolor{red}{1.05} \\
\hline
\end{tabular}
}
\end{table}

\begin{table}[ht]
\centering
\label{tab:fusion_ablation}
\resizebox{0.5\textwidth}{!}{
\begin{tabular}{llcc}
\hline
 Dataset&\textbf{Configuration} & \textbf{ WF1 (\%)}& \textbf{ Acc (\%)}\\
\hline
 \multirow{4}{*}{\textbf{IEMOCAP}}&Full model & 73.35 & 73.10 \\
 &C1: w/o CAF (only multi-head self-attention) & 72.10 $\downarrow$\textcolor{red}{1.25}& 72.00 $\downarrow$\textcolor{red}{1.10}\\
 &C2: w/o gating (concatenation + FC) & 72.40 $\downarrow$\textcolor{red}{0.95}& 72.50 $\downarrow$\textcolor{red}{0.60}\\
 &D1: single fusion  & 72.80 $\downarrow$\textcolor{red}{0.55}& 72.90 $\downarrow$\textcolor{red}{0.20}\\
\hline
\end{tabular}
}
\end{table}

To explore the impact of different information interactions and fusion mechanisms on the performance of multimodal emotion recognition, we designed several ablation configurations based on a complete model that incorporates cross-attention fusion (CAF), GRU-style gating mechanisms, and multiple iterative fusion strategies. Specifically, configuration $C_1$ removes the CAF module and replaces it with a conventional multi-head self-attention mechanism, such that all modalities perform only intra-modal self-attention without engaging in cross-modal weight learning. This setting is intended to verify the effectiveness of cross-modal cross-attention in information selection and reweighting.
Configuration $C_2$ eliminates the GRU-style gating mechanism and substitutes all fused gating units with a ``direct concatenation + fully connected layer'' structure, thereby removing any gating-based filtering operations. This design allows us to evaluate the contribution of the gating mechanism to information screening and memory retention.
Configuration $D_1$ adopts a single-step fusion strategy, wherein the output is produced directly after one application of the ``cross-graph/cross-modal fusion (CAF + gating)'' process, without further iterative refinement. This configuration is designed to examine the difference between single and multiple iterative fusion steps in facilitating fine-grained information interactions.

In the $C_1$ configuration, where the cross-attention fusion (CAF) module was removed, the weighted F1 score (WF1) of the model on the MELD dataset decreased by 1.60\%, and on the IEMOCAP dataset by 1.25\%. These results indicate that the cross-modal cross-attention mechanism can effectively capture dependencies between key modalities and significantly enhance the model’s capacity for information reweighting. 
For the $C_2$ configuration, in which the GRU-style gating mechanism was replaced by a simple concatenation followed by a fully connected layer, the model performance dropped by approximately 1 percentage point across both datasets. This demonstrates the critical role of the gating mechanism in filtering out redundant information and preserving long-range dependencies during fusion.
The $D_1$ configuration, which employed only a single fusion step rather than multiple iterations, exhibited a slight performance degradation compared to the complete model. This observation further confirms that iterative fusion strategies promote more fine-grained interactions among multimodal features, leading to additional performance improvements.
Overall, this ablation study quantifies the contributions of various fusion components and provides valuable insights for the lightweight design and performance optimization of future multimodal models.
\begin{figure*}[!t]
    \centering
    \includegraphics[width=\linewidth]{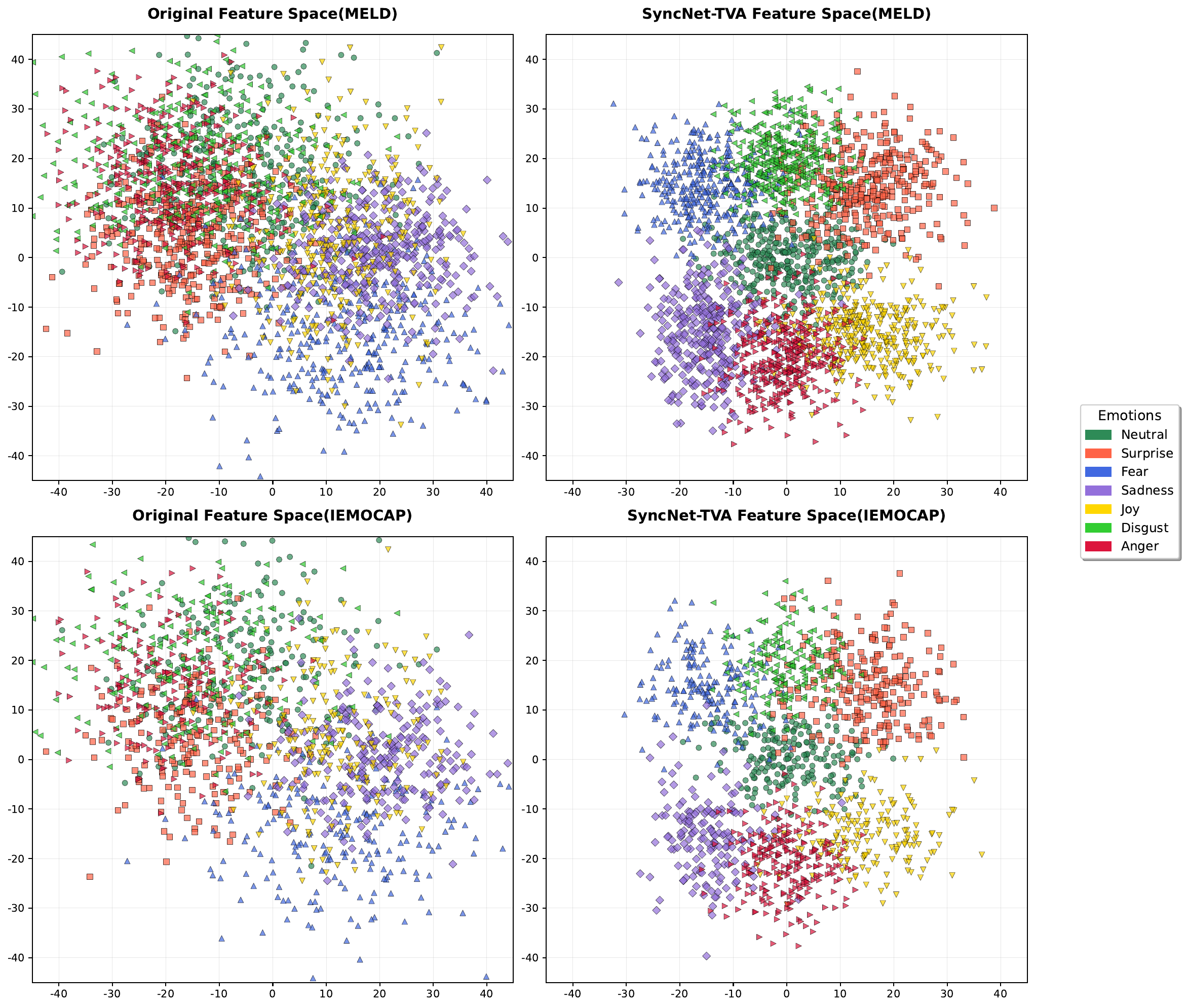}
    \caption{Visualization of feature space before and after processing by Sync-TVA on MELD and IEMOCAP datasets.}
    \label{fig:feature-space}
\end{figure*}

\subsection{Confusion Matrix Analysis}

To better understand the per-class performance of our model, we visualize the confusion matrices for the MELD and IEMOCAP datasets in Figure X. On the MELD dataset, the model shows strong accuracy in identifying Neutral (81.2\%), Joy (66.7\%), and Anger (55.1\%), suggesting that these emotions are relatively easier to distinguish in multi-party conversations. However, emotions such as Fear, Sadness, and Disgust are frequently misclassified, indicating challenges in recognizing subtle and overlapping negative emotions. In particular, Fear is often confused with Sadness (21.5\%) and Disgust (15.1\%), which may be due to limited distinct cues in textual or acoustic features.

In contrast, the confusion matrix for IEMOCAP demonstrates significantly better emotion discrimination. The model achieves high classification performance across most emotions, including Sadness (84.3\%), Excited (80.1\%), Anger (72.2\%), and Neutral (71.5\%). These results highlight the model’s robustness in dyadic conversations with rich multimodal information. Nonetheless, a certain level of confusion is still observed in Frustrated, which is occasionally misclassified as Neutral (6.9\%) or Anger (9.7\%), possibly due to the ambiguous nature of frustration in real-world interactions.

\subsection{VISUALIZATION}

In order to demonstrate the effectiveness of our model in the feature space, we will visually present the original features as well as the features processed by our Sync-TVA model. These images show the features of test samples from the IEMOCAP dataset and the MELD dataset, projected into a two-dimensional space. From the visualization results, we observed that the Sync-TVA model successfully transformed the raw features. The subplots in the upper left and lower left corners show the original features before processing, while the subplots in the upper right and lower right corners display the features after being processed by Sync-TVA. As shown in Figure 3, in the original feature space, samples of different emotion categories (neutral, surprise, etc.) are relatively scattered and mixed with each other; whereas in the feature space processed by Sync-TVA, samples of the same emotion category cluster more closely, and the distinguishability between different emotion categories is enhanced. This indicates that our Sync-TVA model captures more discriminative features for emotion recognition, effectively optimizing the feature distribution, which helps to better distinguish between different emotions.

\section{CONCLUSION}
This study presents a novel multimodal emotion recognition model Sync-TVA, designed to enhance the expressive representation of multi-source emotional information and improve cross-modal collaborative understanding. This model finely models the semantic features within each modality through the 'Modal Specific Dynamic Enhancement Module (MSDE)' and constructs three types of heterogeneous graphs-text-visual, visual-audio, and audio-text-in the 'Enhanced Graph Structure Modeling Mechanism', thus achieving a structured representation of multimodal interaction relationships. On this basis, we introduce the 'Deep Information Interaction Fusion Mechanism', utilizing cross-attention and the Gate Fusion Module (GFM) to achieve deep semantic alignment and supplementation between modalities, effectively enhancing the accuracy and generalization capability of emotion classification. 

Experiments on the authoritative multimodal emotion recognition datasets MELD and IEMOCAP show that Sync-TVA outperforms various existing advanced methods in terms of weighted F1 socre and overall recognition accuracy, especially demonstrating higher robustness and distinguishability in recognizing minority emotions such as 'fear' and 'disgust'. Ablation experiments further confimed the significant improvement in overall performance brought by the MSDE module, graph structure design, and fusion strategy, providing structural empirical support for further modal optimization.
Future research can further develop in three directions: first, introducing a multi-turn dialogue context modeling mechanism to enhance the model's ability to track emotional evolution trends; second, combining contrastive learning or self-supervised pre-training strategies to mitigate training bias caused by class imbalance; third, designing a more adaptive lightweight cross-modal fusion structure to address the issues of noise interference and data loss encountered in practical applications.

\bibliographystyle{IEEEtran}
\bibliography{main}

\newpage

\section{Biography Section}

\begin{IEEEbiography}[{\includegraphics[width=1in,height=1.25in,clip,keepaspectratio]{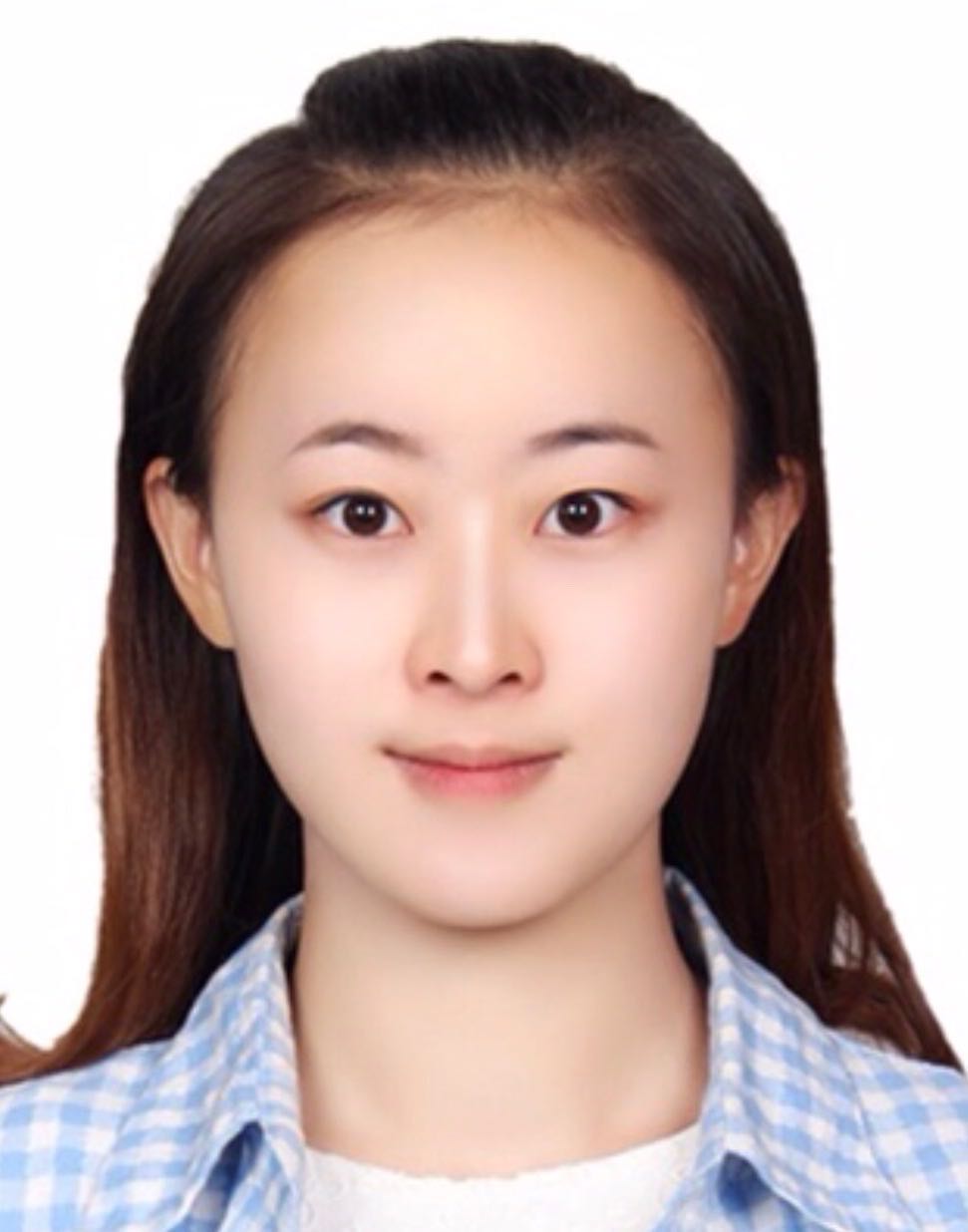}}]{Zeyu Deng}
received her MSc in Engineering and Education from University College London (UCL). She is currently pursuing a PhD in the James Watt School of Engineering at the University of Glasgow. Her research focuses on multimodal perception, multimodal emotion recognition, human-Robot interaction, and swarm robotics.
\end{IEEEbiography}

\vspace{11pt}

\begin{IEEEbiography}
[{\includegraphics[width=1in,height=1.25in,clip,keepaspectratio]{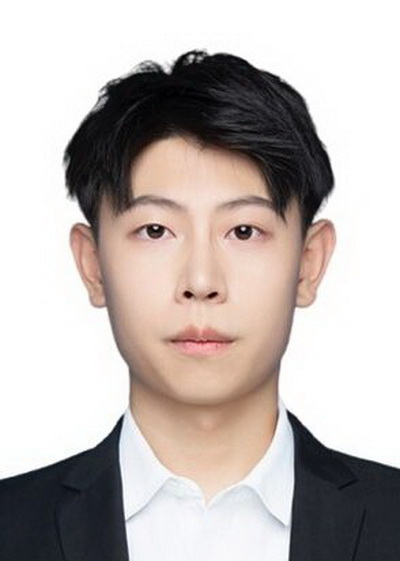}}]{Yanhui Lu}
received his MSc in Robotics from the University of Bristol, U.K. He is currently pursuing a PhD candidate in the School of Engineering Mathematics and Technology at the University of Bristol. His research interests include robotic tactile sensing, multimodal perception, neuromorphic computing, and assistive robotics.
\end{IEEEbiography}

\begin{IEEEbiography}
[{\includegraphics[width=1in,height=2in,clip,keepaspectratio]{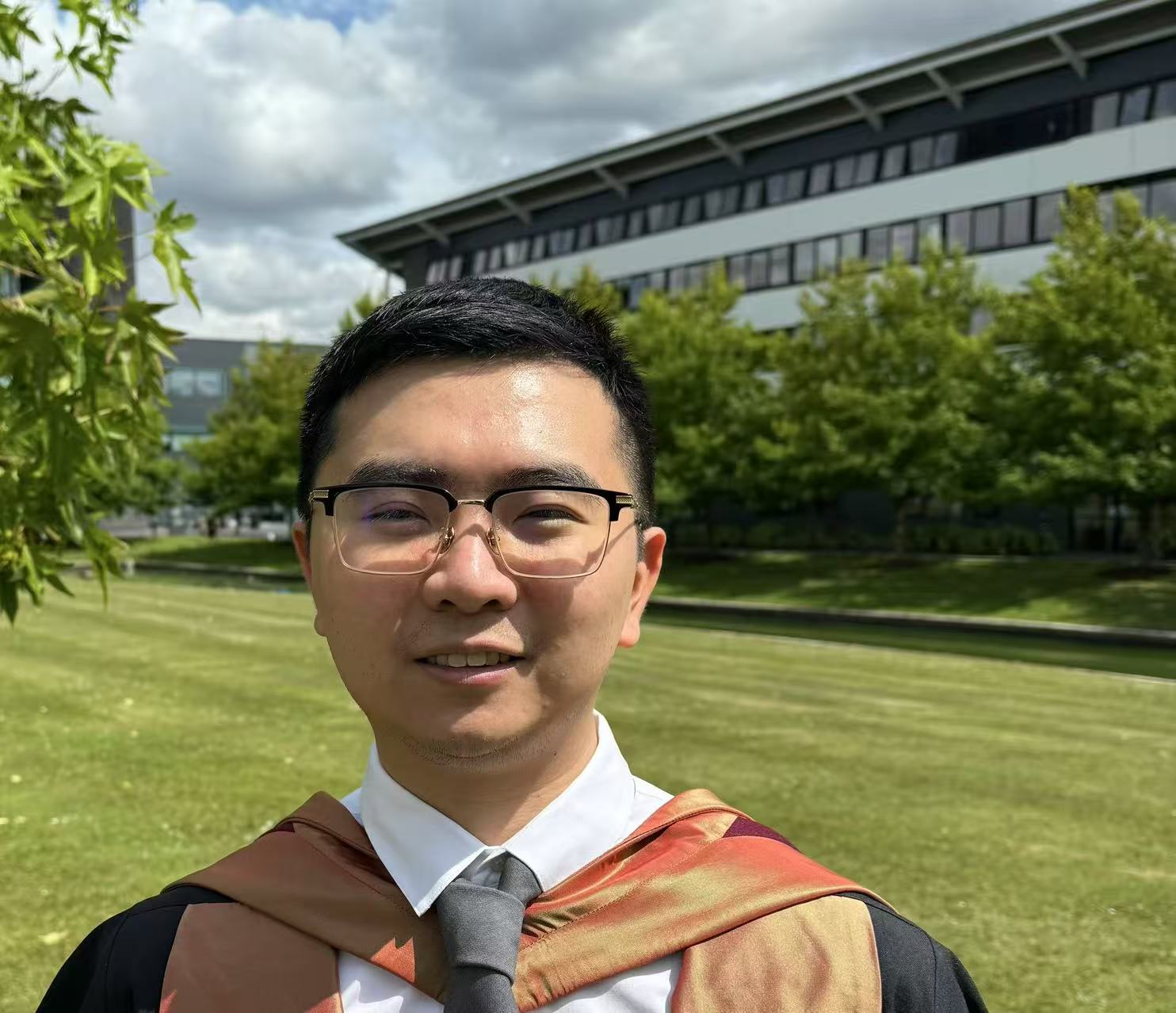}}]{Jiaoshu Liao}
received his Ph.D. in Computer Science from the University of Warwick, U.K. He is currently a postdoctoral researcher at the University of Glasgow. His research interests include image recognition, image segmentation, image generation, and video analysis.
\end{IEEEbiography}

\vspace{11pt}

\begin{IEEEbiography}
[{\includegraphics[width=1in,height=1.25in,clip,keepaspectratio]{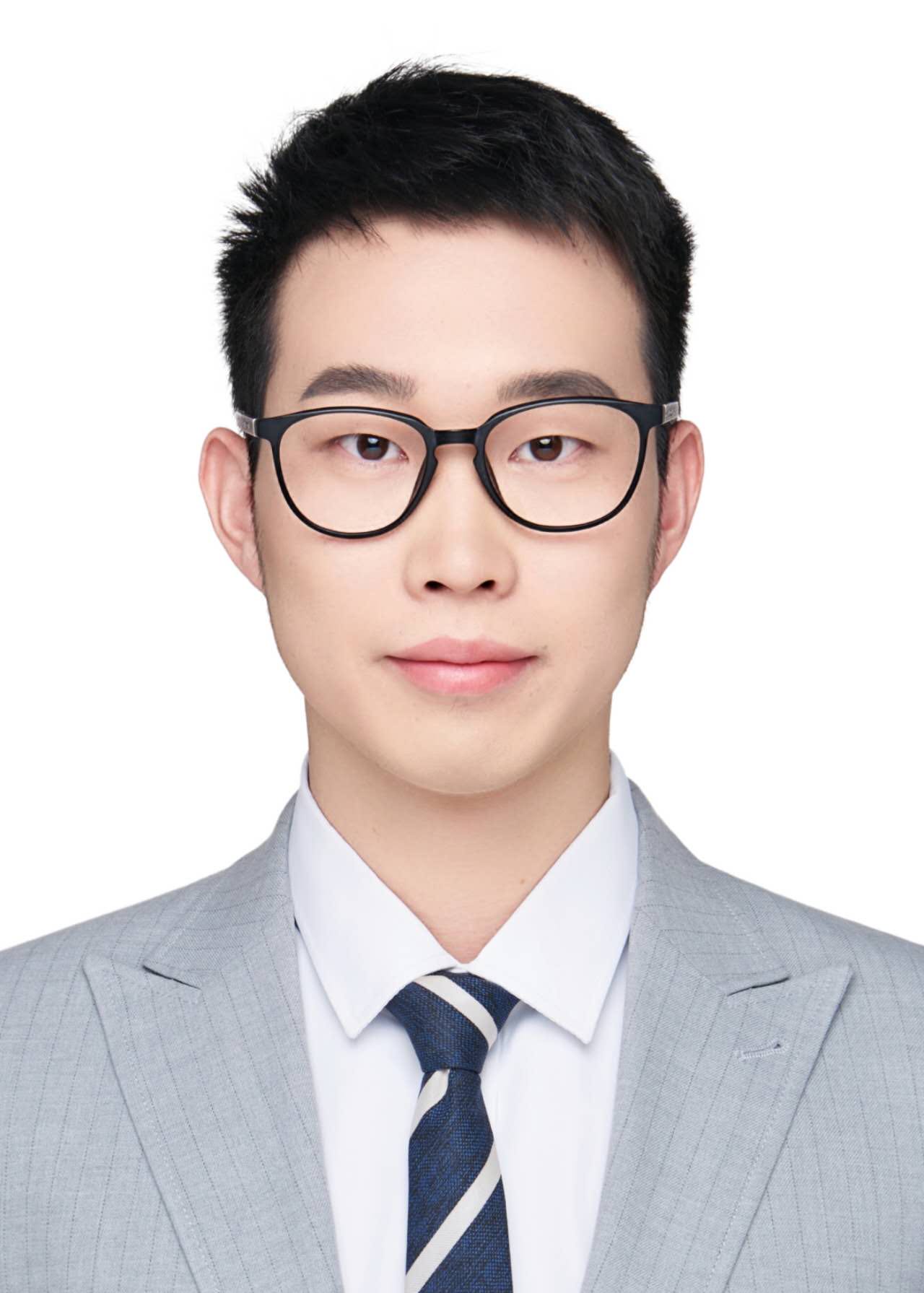}}]{Shuang Wu}
received an MSc in Computer Systems Engineering from the University of Glasgow (UK) in 2022 and is currently a PhD candidate in the Department of Civil, Environmental \& Geomatic Engineering at University College London (UCL), focusing on multimodal emotion recognition.
\end{IEEEbiography}

\vspace{11pt}

\begin{IEEEbiography}
[{\includegraphics[width=1in,height=1.25in,clip,keepaspectratio]{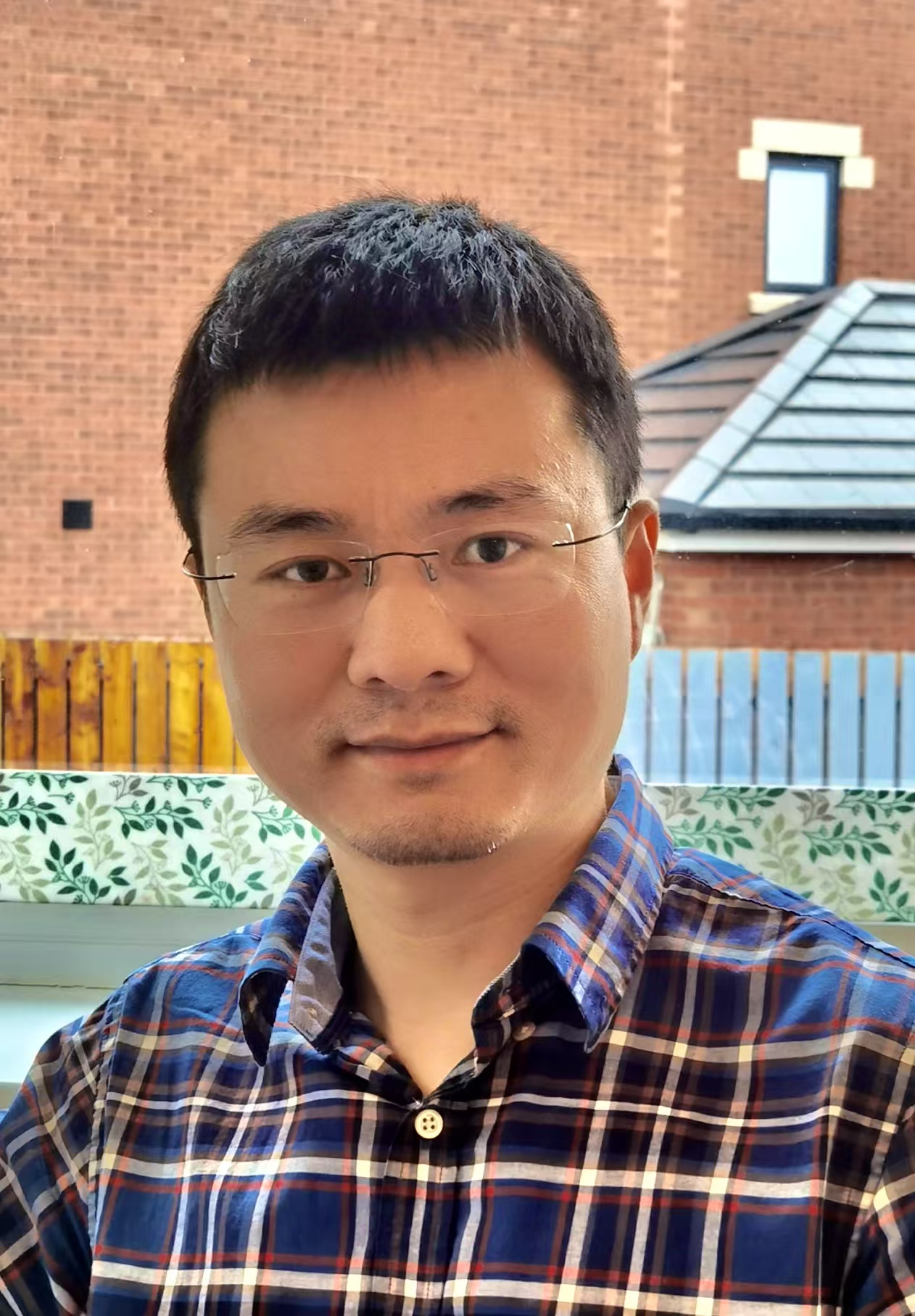}}]{Chongfeng Wei}
received his Ph.D. degree in mechanical engineering from the University of Birmingham in 2015. He is now an Associate Professor \(University Senior Lecturer\) at University of Glasgow, UK. His current research interests include decision-making and control of intelligent vehicles, human-centric autonomous driving, cooperative automation, and dynamics and control of mechanical systems. He is also serving as an Associate Editor of IEEE TITS, IEEE TIV, IEEE TVT, and Frontier on Robotics and AI.
\end{IEEEbiography}

\vfill

\end{document}